\documentclass[preprint,12pt]{elsarticle}

\usepackage{graphicx}
\usepackage{amssymb}

\usepackage{lineno}

\usepackage{graphicx,times,amsmath} 
\usepackage{csquotes}
\usepackage{url} 
\usepackage{float}

\newcommand{\specialcell}[2][c]{%
  \begin{tabular}[#1]{@{}c@{}}#2\end{tabular}}

\journal{-}

\begin{document}

\begin{frontmatter}


\title{Hunting for Spammers: Detecting Evolved Spammers on Twitter}



\author{Nour El-Mawass$^{1}$ and  Saad Alaboodi$^{2}$}

\address{College of Computer and Information Sciences, King Saud University (Riyadh, Saudi Arabia)
\tt\small \(^1\)nour.elmawass at ieee.org, \(^2\)salaboodi at ksu.edu.sa
}

\begin{abstract}
Once an email problem, spam has nowadays branched into new territories with disruptive effects. In particular, spam has established itself over the recent years as a ubiquitous, annoying, and sometimes threatening aspect of online social networks. Due to its prevalent existence, many works have tackled spam on Twitter from different angles. Spam is, however, a moving target. The new generation of spammers on Twitter has evolved into online creatures that are not easily recognizable by old detection systems. With the strong tangled spamming community, automatic tweeting scripts, and the ability to massively create Twitter accounts with a negligible cost, spam on Twitter is becoming smarter, fuzzier and harder to detect.  Our own analysis of spam content on Arabic trending hashtags in Saudi Arabia results in an estimate of about three quarters of the total generated content. This alarming rate makes the development of adaptive spam detection techniques a very real and pressing need. In this paper, we analyze the spam content of trending hashtags on Saudi Twitter, and assess the performance of previous spam detection systems on our recently gathered dataset. Due to the escalating manipulation that characterizes newer spamming accounts, simple manual labeling currently leads to inaccurate results. In order to get reliable ground-truth data, we propose an updated manual classification algorithm that avoids the deficiencies of older manual approaches. We also adapt the previously proposed features to respond to spammers evading techniques, and use these features to build a new data-driven detection system.
\end{abstract}


\end{frontmatter}


\section{Introduction}
\label{S:1}
The rise of online social networks has marked the beginning of a new spam era. Whether it is on Facebook, Twitter or other similar platforms, spam on online social networks is characterized by its invasiveness, ubiquity, and interference with the website's main functionalities. The spammers' aggressive abuse of online social networks threatens of reducing the value and credibility of these platforms, which may destroy their premise of an alternative, secure, and enjoyable online existence.

The ubiquity of spam can greatly degrade the experience of the social network user, replacing a human-human communication with an experience where profit-oriented accounts, mostly bots, pollute the online sphere with insignificant, context-less and even harmful content.

In our investigation, we found that about three quarters of the tweets in trending hashtags in Saudi Arabia are spam messages. A deeper analysis shows that the percentage of tweets generated automatically is even higher. This not only means that Twitter's resources are being consumed by malicious accounts rather than the intended users, but it also implies the need to doubt any statistics or opinion mining results based on these trends. In particular, it becomes legitimate to question the reports that rank Saudi Arabia as the arab nation with the highest number of active Twitter users \cite{arabnews2014}, or that show a \enquote{booming usage} with a penetration higher than \(51\%\) \cite{globalwebindex2014}.

While the first generation of spammers on Twitter was generally naive and had obvious characteristics that helped separate it from the rest of the population, spammers nowadays recur to cheap automated techniques to gain trust and credibility and go unnoticed in the larger crowd. Unlike emails spammers, spammers on online social networks platforms have many evading dimensions, including the most prominent feature that helped single them out in the near past: their social network. 

In this work, we study the contemporary population of spammers on Twitter, specifically those that tweet in Arabic and that mainly hijack Saudi trends, and target the arabic speaking population. We develop new detection features that adapt to the current evading techniques, and we use these features to train a Machine Learning based system that has the goal of detecting spammers. We show that our approach outperforms older state-of-the-art approaches, and is therefore more adapted to the current detection needs.

\subsection{Contributions and outline}
We summarize the contributions of this work in the following points:
\begin{itemize}
\item We take a fresh look on an old methodology applied extensively to Twitter. We reuse the statistical supervised classification methodology and revise all assumptions and steps in it. We follow a data-driven approach, and choose to base our changes on empirical evaluation of the data rather than hypothetical reasoning.
\item We provide unique statistics on the prevalence of spam and automation in trending hashtags written in arabic. Previous work has mainly targeted english and chinese spam \cite{yu2012artificial,zhu2012discovering}, and we hope that this work will contribute to a greater understanding of non-english spam. 
\item We propose an updated manual classification algorithm that avoids the deficiencies of older approaches and takes into account the automation status of the account in question. 
\item We evaluate the performance of three - highly cited - state of the art spam detection systems on recent data. 
\item We evaluate previously proposed detection features, and assess the impact of evasion techniques on these features.
\item We Propose a set of features that are most suitable to the detection of the current Twitter's spammers population.
\item We propose a “hunter” unit that works in adjunction to our main classification unit. This special unit is used to increase the probability of having spammers in the sample. It takes advantage of the social network of spammers that the system has already detected.
\end{itemize}

The rest of this paper is organized as follows. In part II we cover the background on spammers detection on Twitter, we start by introducing Twitter and covering related work on spam detection. We then proceed to the discussion of how to define spam and spammers. Part III is dedicated to data crawling and labeling. In part IV we explain in detail how the features are computed and how they respond to different evasion techniques. The architecture of the system is discussed in part V. In part VI, we show the results and performance of our classifier and compare it to the performance of other state-of-the-art systems. We conclude the paper in part VI, where we also discuss future research directions. 


\section{Background and Related Work}
\subsection{The Twitter social network}
Twitter is an online social network and microblogging platform allowing its users to write and read tweets, which are texts limited by 140 characters that can optionally contain a URL or an image. The tweets of a user appear on his personal page, and in the news feed of his followers. Accounts are public by default, meaning that anyone accessing a user's personal page can see his tweets feed. A user can choose to make his account protected, in which case only his followers can see his feed.

Unlike Facebook's friendship relation, following on Twitter is not necessarily bidirectional: A user \(i\) can follow user \(j\) without \(j\) following him back. In this scenario, \(i\) is a \enquote{follower} of \(j\) and \(j\) is a \enquote{friend} (or alternatively a \enquote{followee}) of \(i\).

A user \(i\) can mention another user \(j\) (not necessarily a friend of \(i\)), by including its user screen name preceded by the '@' symbol in the tweet. The same convention also appears in any reply to a tweet by \(j\).

Interaction with other accounts' tweets has mainly two channels: the first is \enquote{favouriting} the tweet (loosely equivalent to liking it on other online social network platforms)\footnote{Recently, Twitter has renamed \enquote{favourite} to \enquote{like}, but we will be using the old terminology throughout the paper since the work was done before the change took place.}. The second interaction method is known as \enquote{retweeting}. Retweeting means sharing the tweet on the user's own page, while keeping the reference to the tweet's original writer. 

A hashtag is a special text entity that may be formed by one or several words preceded by the \enquote{\(\#\)} symbol. Including a special hashtag in a tweet means linking it to all the tweets that contain the same hashtag. This allows communities to grow around hashtags, and allows the discussion and follow up of specific events by searching Twitter for all tweets containing the relevant hashtag. When a given hashtag gains a lot of popularity and activity in a given region, it becomes a \enquote{trend}, and will show on the Twitter page of users in that region.

Since the number of characters in a tweet is limited, URLs shortening was a common practice in the early days of Twitter. This has the advantage of saving space, but also prevents a user from suspecting that a link is malicious before clicking on it. Nowadays, links are automatically shortened by Twitter, but the practice is still used by some users, mostly to hide the original URL \cite{thomas2011design}.

Manual ways to deal with unwanted attention from some accounts, or to report malicious activity, include blocking an account and reporting it.

\subsection{Related Work}
Previous work on spam detection on Twitter can be typically divided into descriptive and reactive approaches. Reactive work can be further divided based on the used methodology. Our work is built on the machine learning supervised classification model. The same methodology has been previously used in many studies to detect spammers in the population of normal OSN accounts. The work flow is pretty unified and can be summarized as follows: gathering and analyzing a dataset representing the OSN accounts population, creating ground-truth data using a sample of the initial dataset, heuristically defining and selecting appropriate classification features, building a statistical classifier, and assessing the performance of the classifier on the manually labeled dataset, and optionally on a larger non-labeled dataset.

Details of the approach differ and there exist different methods to collect the data. One method is to attract spammers using social honeypots. These are accounts created by the researchers to mimic the average user of the studied social network \cite{stringhini2010detecting,lee2010uncovering,lee2011seven}. This approach requires a long duration (spanning months) of inactive observation before a satisfactory users database is built. The resulting spammers dataset is often biased, as it only comprises spammers that are actively following other users. Following other users in the hope of being seen or followed back is a characteristic of spammers, but it does not apply to the entire spammers population. For example, a sizable portion of the contemporary spammers population ensures visibility by hijacking hashtags and spreading their tweets there. Additionally, the growing followers and retweets selling market allows spamming accounts to \enquote{appear} popular without engaging in a complicated massive following/unfollowing behavior. 

The honeypots method has the additional disadvantage of yielding a dataset that is skewed toward spammers, and does not represent their actual percentage in the larger OSN population. Since normal users rarely follow other users randomly, it is expected that normal users will be under-represented in the gathered dataset, therefore requiring additional sampling of the OSN to equilibrate the dataset. 

A more active approach is to collect users directly either by brute force \cite{benevenuto2010detecting} or by sampling the Twitter sphere. The latter approach involves choosing a small sample of users as a seed, and expanding the network by following the social networks of these users up to a given degree \cite{yardi2009detecting}. Yet another fast collection method is to sample accounts that are tweeting in Twitter's public timeline \cite{wang2010don,stringhini2010detecting} (now discontinued), or in trending hashtags \cite{yardi2009detecting}. Another efficient method is to use Twitter's streaming API \cite{twitterStreaming} to collect a large dataset of tweets \cite{grier2010spam}. 

Compared to other collection methods, social honeypots have the advantage of continuously collecting up-to-date evidence on content polluters from the social network without the need for a human inspecting the gathered data \cite{lee2011seven}. One can argue against this point of view, however, using the observation that benign users may follow honeypot accounts for non-malicious reasons such as the desire of being followed back, or a genuine desire to connect with the account \cite{stringhini2010detecting}. Thus, users following honeypots may well require human inspection of their profiles before being considered malicious accounts.

How the ground-truth data is labeled is another issue. To complicate things further, the research does not agree on a unified definition of a spam tweet, or a spammer. Some approaches infer the spamming status of an account by assessing only one tweet of that account \cite{benevenuto2010detecting}, others require that a given percentage of the tweets (such as 10\%) are spam tweets \cite{yang2011free}, and some do not give an exact threshold to exactly reproduce their chosen approach, but the total number of examined tweets per account is provided, e.g. 20 \cite{wang2011social}.

What constitutes a spam message is another complexity dimension. Is it a tweet containing a malicious URL? If an account repeats the same tweet a given amount of times, is this tweet considered a spam? What are the conditions to consider that an advertisement tweet is a spam?

Some approaches have assessed the safety or suspiciousness of URLs in tweets as a mean to detect spam tweets \cite{thomas2011design,cao2015detecting,wang2013click,lee2012warningbird}. Although rigorous methods were used to build these detection systems, tweets containing URLs form a limited fraction of spam tweets, and containing a safe URL does not mean the tweet is a legitimate, non spam message. This will be discussed further in section 2.3.

More recent work has investigated the relationship between automation and spamming. In \cite{amleshwaram2013cats} for example, a system for “automated spammers” detection is described. Features related to automation have been exploited to adapt to the changing structure of Twitter’s spammers population \cite{yang2011free}. An analysis of automated activity on Twitter is presented on \cite{chu2010tweeting}, and a system that detects the automation of an account is described in \cite{chu2012detectingautomation}.

In \cite{yang2012analyzing}, authors analyze the spammers' social network on Twitter. 

Spam detection has also been studied on other social networks, such as online video social networks \cite{benevenuto2009detecting}, social bookmarking sites \cite{markines2009social}, location-based social networks \cite{costa2014pollution} and other online social networks e.g. Facebook \cite{chu2012detectingsocial}.

The dominating methodology in the work we discussed so far is the supervised classification methodology. Other approaches require little human expert involvement and use graph and time based parameters to cluster and detect malicious activity on social networks \cite{beutel2013copycatch,jiang2015general}.   

\subsection{Defining spam tweets and spammers accounts}
It is easy to get overwhelmed by the variety of automated activity on Twitter. And although most of this activity is malicious, it cannot be directly associated with spam. There is a disagreement on what constitutes a spam tweet or a spam account in the research community, and different papers focus on different aspects of spam. Twitter in particular, has an extensive definition of what defines a spammer in its Terms of Service \cite{twitterToS}. This extensive definition gives Twitter the flexibility of dealing with what it assesses as malicious activity, but does not give definitive answers on whether a given account that achieves some of the characteristics in Twitter's definition is actually a spammer. In fact, some of the characteristics may well apply to a legitimate user. 

For these reasons, we explicitly give our definition of spam on Twitter. This definition is narrower than Twitter's broad definition, but both more concise and general than the definitions given by earlier work on the subject \cite{benevenuto2009detecting,stringhini2010detecting,wang2010don}. 

Before proceeding to the definition of a spamming account, we give here our definition of a spam tweet. A tweet is considered a spam tweet if it satisfies the following condition:
\begin{enumerate}
\item It contains a hashtag, a mention, a URL or an image.
\item And it is out of context. This can apply to the following overlapping definitions:
\begin{itemize}
\item The tweet topic is not related to the hashtag/trend it contains. The topic can be inferred from the text or the image.
\item The URL redirects to a page not related to the tweet text or the tweet hashtag.
\item The URL redirects to a malicious/phishing site.
\item The tweet is advertising a product or a service by hijacking hashtags or mentioning users in an automated way.
\end{itemize}
\end{enumerate}

In addition to this definition, and as of Twitter’s rule, we consider any tweet advertising a paid retweet/favorite service or selling followers to be a spam as well, regardless of
whether it contains an entity (URL/hashtag/mention) or not.

A spam tweet can be tweeted by an account that is human-operated, bot-operated or a mixture of these two types. We consider any automated account that tweets spam to be a spammer. A more careful examination of the account should be done if it is not totally automated, that is if a fraction of the account tweets is tweeted by a human. It may be the case that the human account has been compromised by one of the  malicious applications it subscribed to (more details in the following sections). 

Unlike previous approaches that assessed a tweet by the \enquote{safety} of links it contains (using Google Safe Browsing (GSB) for example) \cite{GSB}, we do not consider the URL to be a vital part of a spam tweet. We back this choice with empirical evidence: \(79.6\%\) of spam messages in our dataset do not contain a URL, yet they still exhibit the same degree of content pollution of a spam tweet containing a URL. We also sampled 5000 arabic spam tweets containing URLs, extracted the URLs, followed the redirections of each individual URL and used Google Safe Browsing API to flag malicious final webpages. Surprisingly, none of the URLs in our sample was flagged as malicious. This could have many explanations, namely that none of the webpages exhibited a phishing or malware spreading behavior, or that the GSB API database did not contain yet the weblinks of suspected phishing or malware pages at the time of our inspection. Regardless of the cause, these empirical findings suggest that the classic definition of a spam tweet as a tweet containing a URL that would be flagged by safe browsing services provides a very skewed view of the spammers scenery on Twitter. Only a non-representative, tiny fraction of the spammers population can be detected this way, therefore severely biasing any further analysis or generalization.

\begin{figure}[h]
\includegraphics[scale=0.5]{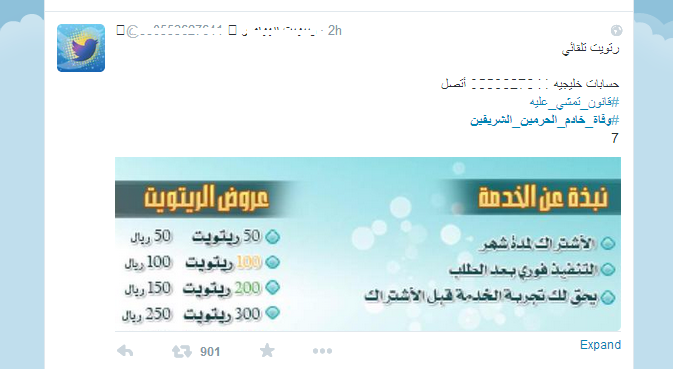}
\centering
\caption{A spam tweet that promotes a followers selling service by hijacking two unrelated trending hashtags. The tweet text says: '[This is an] automatic retweet. Khaliji (Gulf-based) accounts. Call [phone number]'. Note the high number of retweets and the random number added to the end of the text.}
\label{fig:spam_example}
\end{figure}

In fact, we would like to argue that, just like email spam that does not contain actionable content\footnote{A possible reason is detecting what emails will bounce and inferring what email accounts are still active.}, a spam message on Twitter can have motivations other than having users click on a malicious link. An account that sells bags for example, would include the image of its product in the spam tweet and let interested users click on its username to be redirected to its account and eventually read its \enquote{about me} section containing its contact information. This can also lead to users replying to the tweet or privately sending a message to the seller account. The same model applies to a variety of opportunistic accounts on Twitter. These include accounts that sell followers, retweets, products and services. This new perspective on tweets misuse offers a deeper and more realistic angle of the current forms and goals of arabic spam on Twitter.   Fig.~\ref{fig:spam_example} shows an example of a spam tweet, which is misusing two unrelated hashtags. Its goal is to promote a followers selling service, and it does not include any URL in the tweet text.

\section{Data Collection}
\subsection{Crawling Twitter}
We used two main ways to collect our dataset. First, we used Twitter API to crawl 7 trending hashtags in Saudi Arabia in the period between 26 November and 15 December 2014. This resulted in a dataset containing 319,390 unique tweets, and 102,131 unique account identifiers. In addition, we used the streaming API \cite{twitterStreaming} on December 13, 2014 from 12:00 noon until 12:00 night UTC/GMT time, and obtained 22,771,358 unique tweets generated by 1,816,668 unique account identifiers. Table~\ref{table:dataset_info} summarizes the information of the crawled dataset.

\begin{table*}[ht]

\begin{center}
\resizebox{\textwidth}{!}{
\begin{tabular}{|c||c||c||c||c|}
\hline
Source & Period & Nb. of unique tweets & Nb. of unique account IDs
\\
\hline
7 trending hashtags in KSA
& 26 Nov - 15 Dec 2014 & 319,390 & 102,131 \\
\hline
Twitter Streaming API
& Dec 13 2014 (12:00 noon - 12:00 night UTC) & 22,771,358 & 1,816,668 \\
\hline
\end{tabular}
}
\end{center}
\caption{Crawled dataset information}
\label{table:dataset_info}
\end{table*}

\subsection{Building the labeled dataset}
To ensure that the labeled dataset contains a wealth of spammers (both in numbers and variation), we had to recur to a source that is known to attract spamming activity. We selected a trending hashtag related to an important sport event (the 22nd Arabian Gulf Cup final match) and pulled all the tweets that contained the chosen related hashtag (see table~\ref{table:hashtag_stats} for details). 

\begin{figure}[ht]
\includegraphics[scale=0.5]{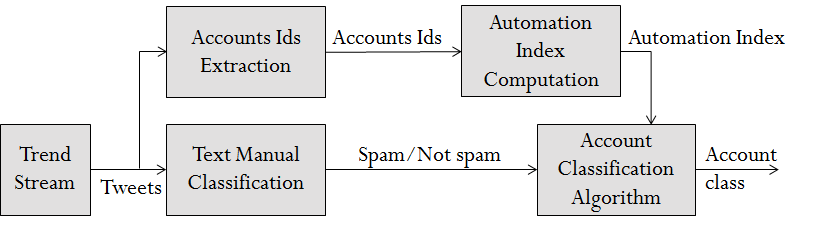}
\centering
\caption{Diagram of labeled dataset building.}
\label{fig:labeled_dataset_building_diagram}
\end{figure}

We randomly chose \(10\%\) of the 55,239 obtained tweets, and classified them by mean of manual examination. \(76.3\%\) of the texts were classified as spam and \(23.7\%\) were classified as legitimate texts relevant to the hashtag context. For each tweet, we pulled the related account IDs (including the original account ID that generated the tweet and the retweeter account ID if existing). We then computed an automation index for each account. Based on a combination of this index and further manual investigation (details to follow), we classified these accounts as spammers, non-spammers, and compromised accounts. The diagram in fig.~\ref{fig:labeled_dataset_building_diagram} summarizes the steps taken to obtain the labeled dataset.

\begin{table*}[ht]
\begin{center}
\resizebox{\textwidth}{!}{
\begin{tabular}{|c||c||c||c||c|}
\hline
Hashtag & Nb. Of tweets & Period & Nb. of classified tweets & \(\%\) of spam tweets
\\
\hline
\specialcell{Who's behind the failure of\\ the saudi football team}
& 55,239 & 26 to 27 Nov 2014 & 5255 & 73.6 \(\%\) \\
\hline
\end{tabular}
}
\end{center}
\caption{Information on the selected trending hashtag}
\label{table:hashtag_stats}
\end{table*}

\subsection{Manual account classification algorithm}
This paragraph is devoted to the description of our manual classification algorithm. During our exploratory analysis of the gathered dataset, we found a need for a detailed classification method that formalizes the ground-truth formation step. This need stems from the fact that results based on simple classification approaches of tweets are often misguiding.

Aside from training datasets based on blacklists, suspended accounts lists \cite{thomas2011suspended}, or safe browsing services \cite{yang2011free}, any approach to create a labeled training dataset requires an amount of manual classification. One direct approach to obtain a dataset where the accounts are classified as spammers/non-spammers is to start from a dataset containing tweet texts labeled as spam/not spam. The next step would be to directly label the account as spammer if the associated tweet text is a spam, and as a non-spammer otherwise \cite{benevenuto2010detecting}. This approach is direct and simple, but the resulting dataset is far from being sufficiently clean. In fact, there are two cases where the direct mapping can go wrong.

The first case is when the tweet text is not a spam: this does not guarantee that the account that generated it is not a spammer. With the heavy masking behavior that spammers use, it is only natural to find a proportion of their tweets to be completely legitimate. 

The other case is when a tweet is a spam, and the related account is not a spammer. The account can be a human account (most tweets are non-automated) that subscribed to a low-traffic app with occasional spamming behavior. We denote this specific case as compromised account.

We first started by examining the source of each tweet in the most recent 200 tweets of the account. The source of the tweet cannot be seen on the user's interface of Twitter, but a developer can access it through API. It is a text field in the pulled description of the tweet that contains two values: a description of the source of the tweet, and a URL pointing to the website that operates the source. The source can be one of Twitter's official sources (such as the web client, the mobile web, an official mobile device application, etc...), a trusted source (such as the Echofon Android application), or a random source that generates automated tweeting activity. Note that the manual classification of the tweets sources into these categories (that we defined) requires an individual understanding of the history and status of each source, and an extensive investigation of its associated URL (if provided). Examining each source is a time-consuming job, and methods to automatically assess the trustworthiness of a source may prove to be useful since a static dictionary will probably become quickly outdated. In fact, we had more than 300 distinct sources in our 300k tweets dataset. 

\begin{figure}[ht]
\includegraphics[scale=0.7]{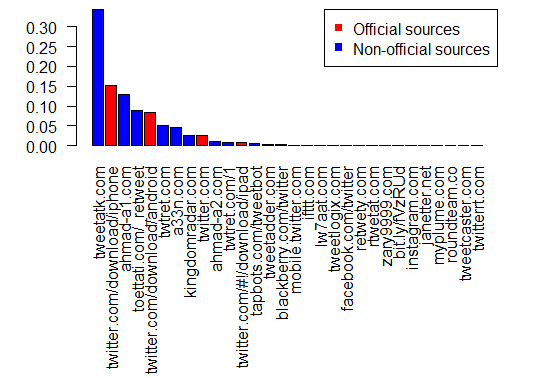}
\centering
\caption{Barplot showing the percentage of content generated by different sources in a trending hashtag containing 3,782 tweets.}
\label{fig:barplot_sources}
\end{figure}

To illustrate the ratio of automation in trending hashtags, we chose a rather small hashtag and plotted the number of tweets of each source appearing in this hashtag (see fig~\ref{fig:barplot_sources}). Despite having a limited number of tweets, this hashtag clearly illustrates the observation that most of the activity is generated by random automated sources rather than human-related sources. More popular hashtags attract a much more aggressive automated activity, and the associated barplot quickly becomes crowded and unintelligible.

Some human-operated accounts subscribe to automated applications resulting in a fraction of their feed being generated by automated sources. We considered an account \enquote{automated} if more than \(80\%\) of its feed was generated by automated sources.

\begin{figure}[ht]
\includegraphics[scale=0.5]{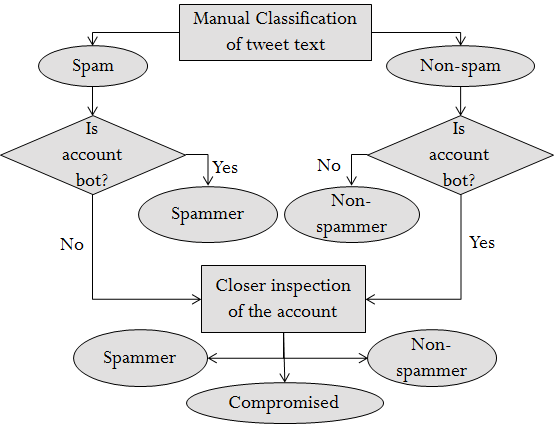}
\centering
\caption{Account classification diagram.}
\label{fig:account_class_algo}
\end{figure}

Fig.~\ref{fig:account_class_algo} summarizes the approach we took to move from our labeled texts dataset to a labeled accounts dataset. The main dogma we followed was two fold and can be summarized as follows: 
\
\begin{enumerate}
\item If the evidence (the labeled text) shows that an automated account is a spammer, take the evidence and consider the automated account to be a spammer. If it shows it as a non spammer investigate the account tweets further before accepting the verdict.
\item If the evidence (the labeled text) shows a human as a non spammer, take the evidence. If it shows it as a spammer, investigate the account's page further to see if the spamming behavior is constant or may just be a slipping or an impulsive retweet.
\end{enumerate}

This method allows us to verify the status of suspicious accounts without having to go manually through all the accounts to ensure the verdict of the labeled texts dataset. Thus it provides a compromise between the accuracy of manually inspecting the accounts (assessing around 200 tweets per account) and the speed of assessing only one tweet (as it has been used in works such as \cite{benevenuto2010detecting,yang2012analyzing}).

\section{Features}

\subsection{Profile attributes}

Using features related to the social network of a profile is a legacy of the earlier work on spam detection. This explains why a lot of evading techniques have the exact aim of evading these features. These features include the number of followers, the number of friends, and the relationship between these two variables, which can be captured using different expressions:
\begin{enumerate}
\item The simple followers per friends ratio: \(\frac{nb(followers)}{nb(friends)}\)
\item an alternative ratio measure \(\frac{nb(friends)}{nb(followers)^2}\)
\item Reputation \(\frac{nb(followers)}{nb(friends)+nb(followers)}\)
\end{enumerate}

In addition to the static numbers, we can also look at these measures in time by computing the number of followers (resp. friends) acquired per day. Since the information on when each relationship was created is not available, we used the age of the account as the time period over which these relationships were created. Hence, the obtained value is the mean value over time: \[nb_{/day}(followers)=\frac{nb_{total}(followers)}{age_{days}(account)}\]
Other general features of the account include the total number of tweets, the number of lists containing the account, the number of tweets favorited by the account, the age of the account, and its tweeting frequency (the average number of tweets generated per day), computed as:
\[\frac{{nb}_{total}(tweets)}{age_{days}(account)}\]

We noticed, however, that some spamming accounts trick both the age and the tweeting frequency features by remaining relatively idle for an extended amount of time after the creation of the account. ٍSimilarly, recently compromised accounts that have been \enquote{normally} active for a long time have a \enquote{clean} history that makes their overall temporal features rather normal. Thus, we introduced the \enquote{recent tweeting frequency}, which computes the frequency over the last 200 tweets only. The former frequency will be denoted as the \enquote{global tweeting frequency}.

\subsection{Content attributes}
Content attributes are based on the recent activity of the profile, namely the last 200 tweets. This specific number is chosen because it is the maximum tweets count that can be obtained using only one API call. Obtaining more than this number would require more API calls, leading to a longer extraction time. Therefore, 200 tweets seems like a rational choice that wisely uses available resources.

That said, spamming accounts created by the more involved spamming campaigns are usually heavily multi-sourced. This means that their tweets are generated by several automatic - possibly independent - tweeting engines. Some of these sources can be legitimate automatic sources that have a low frequency traffic. This type of sources is usually also used by normal non-spamming accounts. The typical content contains quotes, sayings, prayers, jokes, etc... Other types of automatic tweeting scripts that lay on the darker side of the gray spectrum include sources with higher traffic, that are based on custom libraries of thousands of ready-to-use tweets. These sources are often used in conjunction with spamming and advertising sources, usually to mask the spamming content of the account (sybil or compromised) in question.

The most general descriptors of a profile's content are the rates of retweets, replies and of original tweets (they sum to 1). These attributes assess the interaction of the account with its environment and how much of its content is self-generated. On the tweet level, we computed the minimum, maximum, median and average number of words per tweet.
\newline


\subsubsection{Entities-related attributes}
We will use \enquote{entity} to denote the non natural language parts of the tweet, namely hashtags, URLs and mentions. It is natural to expect that the distributions of usage of these entities is different between spammers and non-spammers. We therefore use several features formulations to capture this difference.

On the global level (over the most recent 200 tweets), 3 features can be extracted: the proportion of tweets with URLs, hashtags and mentions, respectively. In addition, since spammers tend to repeat the same URL, hashtag or mention in their tweets for advertisement and visibility reasons, we compute the number of unique entities and the average number of times a unique entity is used. For URLs for example, a normal user is expected to have an average usage of 1 per url, meaning that he used each unique URL only once. A spammer on the other hand tends to have a much higher average usage per unique URL. We use the following expression to compute this value:
\[\frac{nb(\textup{URLs})}{nb(unique(\textup{URLs}))}\]
Note that the unshortened final URL is used, since the shortened URL will be different each time the tweet is generated.

The existence of these entities can be also assessed on the individual tweet level by computing the minimum, maximum, median and average number of occurrences of each type of these entities in a tweet. For example, we can compute the average number of hashtags per tweet (computed by averaging this measure over the most recent 200 tweets). The number of features obtained this way is 12 (\(4 \times 3\)). We also computed the same measures per word per tweet, increasing the number of features to 24.

A technique used by spammers to mask their heavy use of a limited number of entities is to mimic a normal user's behavior by introducing other entities and using them only once. This decreases the average number of uses of an entity, gearing it toward a value close to that of a legitimate user. To account for this behavior, we introduce the \enquote{diversity index}, usually used in ecological and social sciences to measure the adjusted number of species in a population, while taking into account that these species must be evenly distributed. We use the following \enquote{True Diversity} expression to compute the diversity index \(1/\sum_i p_i^2\), where \(p_i\) denotes the rate of usage of an entity \(i\) .

For example, a spammer who has used a hashtag \(h_1\) 147 times, and 20 other hashtags once each, will have the number of unique hashtags equal to 21, while his hashtag diversity index will be close to one (precisely \(1.29\)), thus reflecting the true diversity of hashtags in his tweets. Another example is a legitimate user who used 9 URLs, once each. This user will have his URLs diversity index equal to the number of unique URLs (9), since the probability of usage of each URL is the same. 

Note that this behavior can also be the result of the evolved spamming accounts being multi-sourced. Some of the sources generating the content of the spamming account act like a "white noise", with a content that is closer to normal than to spamming behavior.

We compute the diversity index for URLs, hashtags and mentions. And we use the result to compute an adjusted value of the average number of uses of each of these entities.


\subsubsection{Content replication}
In addition to entities replication discussed in the last section, one of the most prominent features of spamming accounts is the replication of the tweet text itself (possibly with different URL/hashtag/mention each time). Most of the spamming accounts are automated, they can't be as original and random as a human, and unless they have a really diversified spam content, this feature is in essence non-evadable. In practice, however, spammers use a number of evading techniques such as masking the replication with automatically generated legitimate content. 

Another evading technique aims at tricking the similarity measure by adding artificial (meaningless) variation to the tweet text (such as the random three letters word 'ibf'). That's why we use our own definition of similarity based on the Levenshtein edit distance (rather than checking if the two texts are completely identical). To ensure that the similarity is computed on the core text of the tweet and doesn't take into account variables, we filtered the tweet text by removing any hashtag, URL, or mention (including the retweet special identifier 'RT').  Due to these two evasion factors, this feature is not currently as efficient as it was in detecting spammers.

A tweet text is considered a duplicate of another if the similarity is higher than \(90\%\). This threshold misses some duplicates that are carefully twisted (especially if the tweets are semantic duplicates\footnote{We have not encountered semantic duplicates in our arabic tweets dataset. We think the cause is the lack of services similar to Spinbot \cite{spinBot} in arabic.}  rather than syntactic replicates \cite{yang2011free}) but is still able to detect the duplicates that evade the rigid (exact) similarity measure. Techniques to add artificial difference to the tweet include adding a random or iterated number to the tweet (incremented by one each time the same tweet is generated), or adding randomly generated short words (such as xjl, jz1, qic).

The basic features related to this are:
\begin{enumerate}
\item the average similarity computed as:
\[\frac{\sum_{ti,tj \in T, i<j}similarity(t_i,t_j)}{|T||T-1|/2}\]
where T represents the tweets extracted from the account.
\item the number of replicates (the number of tweets tweeted before by the account). 
\end{enumerate}
 
In practice, the average similarity measure is of little help, it does not clearly capture the replication aspect of an account. The number of replicates is relatively robust, since the existence of replicated content cannot be wiped by the the automated masking activity. However, if the account is a high-frequency account, and uses masking abundantly and spamming occasionally, replication can be missed (due to the 200 tweets limitation).
 

\subsubsection{Content reputation}
The reputation of a specific tweet can be assessed using two measures, the number of times it was retweeted, and the number of times it was favorited. A tweet acquires a reputation if its content is relevant to a lot of people, or if the account that generated it is famous/important. When neither of these conditions explain the high reputation of a tweet, this reputation can be alternatively explained by an automated retweeting/favoriting activity, which is generally closely related to spam content. 

To capture tweet reputation we extracted features that measure the statistics over 200 tweets of an account, namely the minimum, maximum, median and average number of retweets and favorites per tweet (8 features in total).

\subsubsection{Spam dictionary}
A feature that is directly inspired by spam detection in emails, measures the proportion of tweets containing a word from a spam dictionary. Since the masking behavior can dramatically decrease the proportion of spam tweets in a spamming account, applying this feature on an account content may not be helpful in detecting complex spamming accounts. 

An arabic dictionary for spam terms is not readily available. In addition, spam on Twitter does not appear to have the same lexicon of emails' spam. Therefore we created a dictionary based on the spam tweets in our labeled dataset. The obtained list was manually filtered, and only the most frequent and relevant words were kept.


\section{Classification}
\subsection{Building a spam classifier}
\begin{figure}[h]
\includegraphics[scale=0.5]{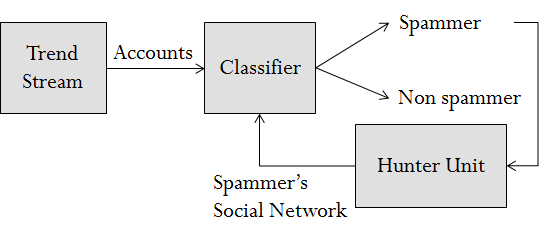}
\centering
\caption{Working system architecture.}
\label{fig:working_system_architecture}
\end{figure}

The targeted classifier is a binary classifier that outputs one of two results: spammer or non-spammer. The core classifier unit in fig.~\ref{fig:working_system_architecture} is fed with the features extracted from the account. A machine learning algorithm decides the output based on the values of the features. And in the case where a spammer is detected, the account details are sent to a \enquote{hunter} unit that extracts the social network of the account, and feeds the obtained accounts back to the classifier.

\subsection{Training methodology and evaluation metrics}
We train several ML algorithms on our labeled data: AdaBoost, Decorate, Random Forest and Naive Bayes. Due to the limited size of the dataset, our best option is to perform a 10-fold cross validation training over the labeled dataset. We assess the performance in terms of the false positive rate, the recall and the F1-measure of the spammers class. 

\subsection{Getting more spammers, the Hunter unit}
A direct consequence of the evasion techniques developed by the spamming community, is the heavy clustering of spammers. To give each other credibility, spammers need to follow each other, and to favorite/retweet the content of their peers. Finding a spammer account is therefore equivalent to finding a mine of spammers. By constructing the social network of a spammer, namely his followers and the accounts that retweet his content, we can add a high-prior list to the accounts dataset fed to the classifier, and increase the overall number of detected spammers.

\subsection{Feature selection}
We use the information gain and the Chi squared selection techniques to select the features to use in our classifier. We obtain the same set of top 20 features from the 2 techniques. After removing correlated features that refer to very similar measures, we are left with the set in fig.~\ref{featuresSet}, to which we add the \enquote{average number of  retweets per tweet} feature.

   \begin{figure}[thpb]
      \centering
      \framebox{\parbox{3in}{
      tweeting frequency per day \\
      proportion of replies\\
      nb. replicates\\
      fraction of tweets with mentions\\
      adjusted nb. of uses of url\\ 
      nb. of hashtags per day\\
      adjusted nb. of uses of hashtag\\
      nb. mentions\\ 
      nb. mentions per day\\
      adjusted nb. of uses of mention\\
      min nb. words per tweet\\
      max nb. mentions per tweet\\
      avg nb. mentions per word in the tweet\\
      + avg nb. retweets per tweet
      
}}
      \caption{The chosen set of features}
      \label{featuresSet}
   \end{figure}

\section{Results}
Using a 10-fold cross validation over our labeled dataset, we obtained the performance of four Machine Learning algorithms in detecting spammers accounts, namely Naive Bayes, Random Forest, AdaBoost and Decorate. We used Weka's implementation of these algorithms, and repeated the process over four different sets of features: Our selected set of features denoted as \enquote{Ours}, the set in \cite{benevenuto2010detecting} denoted as \enquote{Benevenuto}, the set in \cite{stringhini2010detecting} denoted as \enquote{Stringhini}, and the set in \cite{wang2010don} denoted as \enquote{Wang}.

In order to exactly reproduce a classification system described in a previous work, one should possess three key information: 1) the statistical model used to build the classifier, 2) the set of features extracted from the dataset 3) the initial training and testing datasets used to train and test the model (or possibly one dataset with a reproducible way to re-partition it according to the exact authors' method). After obtaining the exact parameters of the model, it can be used thereafter to classify new instances (preferably drawn from current data), and assess the current performance of the model.

Since we did not have access to the initial datasets of the authors in the cited work \cite{stringhini2010detecting,wang2010don} \footnote{The dataset from \cite{benevenuto2010detecting} is available by direct request to the first author. }, we reproduced their work using the different sets of features  described in their respective papers. 

\begin{figure}[thp]
\centering
\includegraphics[scale=0.5]{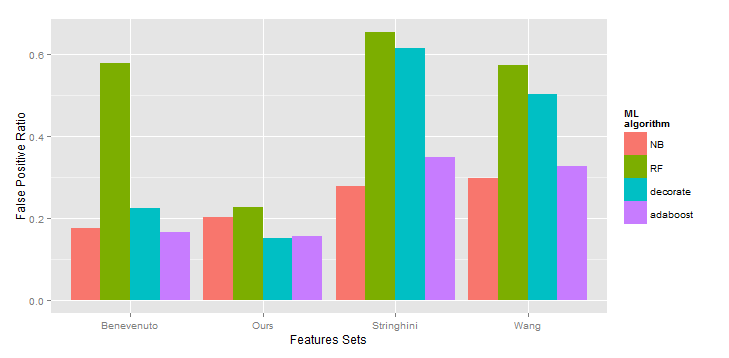}
\caption{False positives rate of the features set using 4 ML algorithms.}
\label{fig:fprate}
\includegraphics[scale=0.5]{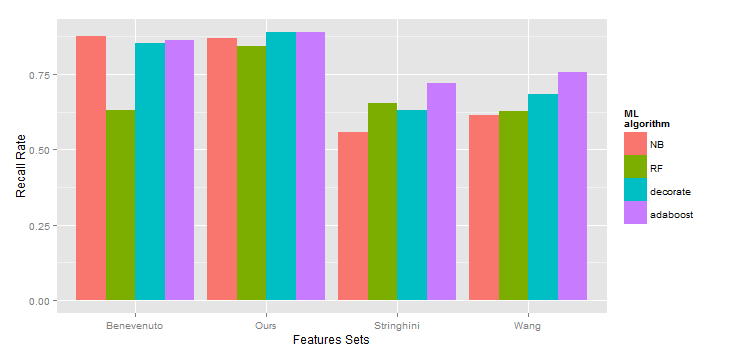}
\caption{Recall rate of the features set using 4 ML algorithms.}
\label{fig:recall}
\end{figure}

\begin{figure}[tp]
\centering
\includegraphics[scale=0.5]{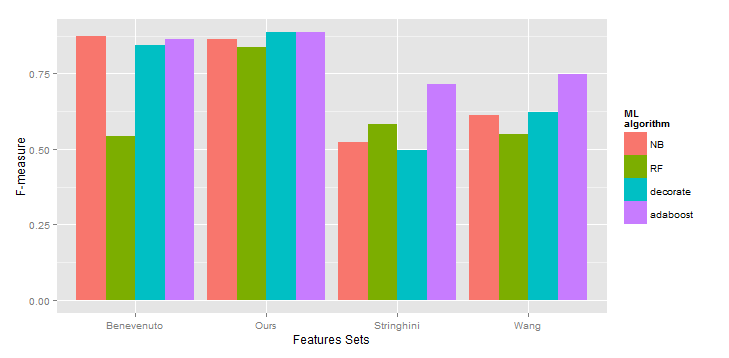}
\caption{F-measure of the features set using 4 ML algorithms.}
\label{fig:f_measure}
\end{figure}

Note that the compared models have been trained over recent data. Therefore, their performance as shown in this manuscript are actually a higher bound of their real performance. We expect that the initial models - with the same sets of features - trained over their original respective datasets would have a significantly worse performance over our current test data. Should we have access to the original datasets, two conclusions could be drawn:
\begin{itemize}
\item The statistical characteristics of the spamming population have significantly shifted.
\item The set of features usable in classifying malicious activity needs to be updated. 
\end{itemize}

These two factors can be independent. The spammers' population can change their behavior, making an old detection system ineffective. Though retraining the model with the same set of features over newer data can adjust the model's parameters and make it effective again.

Our analysis shows that, in drifting away from their old behavioral models, spammers have both changed their statistical description and the features that are effective in catching them.

We report the results over three performance measures following the style in \cite{yang2011free}: the false positives rate (fig.~\ref{fig:fprate}), the recall (fig.~\ref{fig:recall}), and the F-measure (fig.~\ref{fig:f_measure}).

Compared to other features sets, our set and that of Benevenuto \cite{benevenuto2010detecting} show a significantly higher performance in all performance measures. In addition, our set's performance matches or slightly  surpasses that of Benevenuto. This is an expected result due to the high similarity between our set of features selected by Information Gain and Chi squared techniques, and the features set proposed in \cite{benevenuto2010detecting}.

Although some prior work hinted at the importance of including automation related features as a performance booster, we did not obtain a significant improvement to the shown results by including a feature that explicitly states the automation status of the account (at least over our dataset). In other terms, we used our prior manual classification of the accounts as bots or humans, and used the class of each account as an input to the classifier. This did not result in a significant improvement in the accuracy of the presented spam classification model. In fact, it appears that the information brought by the automation feature is already included in the selected set of features that we used. Concretely, in order to build an automation detection classifier, the set of features to be used is highly overlapping with the set of features presented in this paper.

Some important observations concerning previously formulated features (in the literature) can be made. First, none of the naive features related to the social network of an account (namely number of friends/followers and the different ratios combinations) made it to the selected set of features (this can be explained by the evading techniques such as spammers clustering, and followers buying). The same can be said regarding the age (evaded partially by remaining idle for an extended period, or by compromising legitimate accounts), the number of tweets and the similarity features (evaded using automatically generated, legitimate-looking content). Even a rigid replicates measure can be easily evaded using artificial random terms or spinning (the process of using semantic alternatives of a text, which was not observed in our arabic dataset). 

In contrast, some content related features, including those that were first suggested in \cite{benevenuto2010detecting}, are still valid in detecting spammers. Nevertheless, a general observation is that, given the elaborated evasion techniques developed and used by the spamming community, these content features alone are not enough to achieve an adequate accuracy, and new features are therefore needed.
In particular, the used spam dictionary, being static in nature, cannot prove beneficial beyond the time-limited dataset from which it was extracted. Having a spam dictionary that can be effectively used as a detection feature requires building this dictionary in a dynamic and environment-responsive way.


\section{Conclusion and Future Work}

Despite many spam detection works and years of improvement, spam is on the rise in online social networks. Due to the active spamming community and the ever-updating evasion techniques, most of the previously proposed detection methods has become obsolete, and even the best performance comes short of what is needed in a production system.

In this paper, we discussed spamming on Twitter, especially with regard to arabic language spam. We presented a concise definition of spam and spammers, and used it to propose a detailed reproducible manual classification algorithm that takes into account the percentage of automated content in an account. We also presented the evasion techniques developed by spammers and how the features can be adapted to account for the evasion effects. Using these features, we assessed the performance of our approach compared to three previous approaches, and showed that a small subset of features proposed in the literature is still valid. Although these features are usable, the initial statistical model created based on these features needs to be updated using recent data to account for spammers moving away from their previous \enquote{statistical} definition.

Backed with the findings of this work, we think that research on spam detection on Twitter, including ours, has taken a limited approach to feature extraction, and that detection needs to look at spammers as a \textit{community} rather than individual accounts. Additionally, we think that the content of spamming accounts has been barely considered since most approaches focus on detecting spammers as opposed to detecting spam. Detecting spam may turn out to be more rewarding in that a small spam seed will result in a mass detection of accounts that duplicated this seed content \cite{egele2013compa}. Based on this, we plan to investigate the possible extension of this work in three different directions.

(1) The temporal aspect of the tweets: Including a time component in the features may help block evading techniques developed to evade from static features that only look at raw numbers. In fact, an initial analysis showed that spammers have a distinctive temporal pattern that may be either caused by automation, or used to mask automation (depending on the considered observation period and the account in question).

(2) Detecting spam based on text mining techniques and natural language processing: What makes an account a spammer is the tweets content. When a human looks at an account in an attempt to classify it, the main feature in determining his decision is whether he finds an explicit spam content or not. Other features of the account may give hints as to how suspicious the account seems or where to look in the feed, but they do not form a convicting evidence. Previous work has looked at this subject \cite{benevenuto2010detecting,martinez2013detecting,wang2015making,santos2014twitter}, but our preliminary experiments show that there is room for improvement. Aside from older models becoming obsolete, having a spam dictionary is a crucial part of building a system to detect spam text. Yet, relevant spam terms on social networks are constantly changing; thus imposing the need for a more dynamic and time-dependent approach. 

(3) A main difficulty of this work was that accounts cannot be cleanly divided into spammers and non spammers. There is a gray area populated by automated accounts that generate low value content, and by compromised accounts that were created and operated by actual human users before becoming manipulated by spamming applications and campaigns \cite{egele2013compa}. In fact, most of the false positives are accounts that belong to this area. We think that including more than one type of content polluters as possible outputs to the classification systems may help clear up some of the ambiguity, and help Twitter take a different course of action for each type of accounts.



\bibliographystyle{model1-num-names}
\bibliography{spam_twitter.bib}







\end{document}